\documentclass[sigconf,nonacm,anonymous=false]{acmart}

\usepackage{amsmath}
\usepackage{amssymb}
\usepackage{booktabs}
\usepackage{graphicx}
\usepackage{subcaption}
\usepackage{algorithm}
\usepackage{algpseudocode}
\usepackage{xcolor}
\usepackage{pgfplots}
\pgfplotsset{compat=1.18}

\newcommand{\R}{\mathbb{R}}
\newcommand{\e}{\mathbf{e}}
\newcommand{\x}{\mathbf{x}}
\newcommand{\W}{\mathbf{W}}
\newcommand{\bias}{\mathbf{b}}

\setcopyright{none}
\acmConference[RecSys '26]{ACM Conference on Recommender Systems}{2026}{TBD}

\begin{document}

% ==============================================================
% Title and Authors
% ==============================================================
\title[Context Features Are Cheap]{Context Features Are Cheap:\\
Rank-Aware Decomposition for Efficient Feature Interaction in Recommender Systems}

\author{Yevgeny Tkach}
\affiliation{%
  \institution{Taboola}
  \country{Israel}
}
\email{yevgeny.t@taboola.com}

% ==============================================================
% Abstract
% ==============================================================
\begin{abstract}
Modern industrial recommender systems use a deep ranking model to score
$N$ candidates against the same user and context features. Standard
implementations broadcast context features early in the forward pass,
redundantly computing context-only operations $N$ times per request. We
present a rank-aware decomposition applicable to the dominant
interaction mechanisms in modern recommender architectures---Factorization
Machine (FM) pairwise products, Deep Cross Network (DCNv2) cross
layers, self-attention, and fully connected (FC) projection layers---built
on a single algebraic principle: any linear or bilinear operation over
a rank-partitioned input admits an exact block decomposition that moves
context-only computation from once-per-candidate to once-per-request,
identity-equivalent to the original model. Closed-form analysis and
controlled ablation verify that savings scale quadratically with the
number of context features. Applied to a production DLRM-style ranker
without any architectural change, the decomposition increases per-pod
throughput by $87.5\%$ (a $47\%$ reduction in peak pod count) at
identical model predictions.

The identity-equivalent decomposition applies only at the first layer
of cross networks and self-attention, since each layer mixes ranks in
its output. To extend savings across depth, we further introduce rDCN,
an architectural variant of DCNv2 that maintains rank discipline across
depth and matches DCNv2 accuracy within training noise at $67\%$ fewer
total FLOPs, and sketch an analogous architectural variant for self-attention.
\end{abstract}


\maketitle

% ==============================================================
% Body
% ==============================================================
\section{Introduction}
\label{sec:introduction}

Industrial recommender systems are typically organized as a two-stage
retrieval and ranking pipeline. A cheap retrieval stage narrows a corpus of
millions of items down to $N$ candidates per request; a deep ranking model
then scores all $N$ candidates against the same user and context signals
(user profile, browsing history, session state, page context). The ranking
stage faces a fundamental tension between model expressiveness and serving
cost: as models incorporate richer feature interactions---through pairwise
products, cross layers, or attention---the total cost of multi-target
scoring grows with both model depth and the number of candidates evaluated
per request. For production systems serving billions of daily requests, this
cost directly translates into infrastructure footprint.

Most modern ranking architectures---DLRM~\cite{naumov2019dlrm},
DeepFM~\cite{guo2017deepfm}, DCNv2~\cite{wang2021dcnv2}, and their
variants---share a common structural property: they combine context and
target embeddings into a unified feature vector before feeding it through
interaction and projection layers. Many implementations broadcast context
features to the candidate batch dimension before any computation begins. Downstream
operations that involve only context fields---context-only pairwise
interactions, linear projections,
attention---are redundantly computed $N$ times. For a model with $K$ context
fields and $M$ target fields scoring $N$ candidates per request, the
context-context FM interactions alone waste
$(N-1) \cdot \binom{K}{2} \cdot D$ FLOPs per request, where $D$ is the
embedding dimension.

Recent work on serving optimization addresses this redundancy in fully
connected (FC) layers, where the savings scale \emph{linearly} with the
number of context fields. In DLRM-style architectures, the dominant
industrial paradigm, the bulk of feature-dependent computation occurs in
the \emph{interaction} layer, where pairwise combinations cause savings to
scale \emph{quadratically}, $O(K^2)$. As models incorporate richer context
(longer user history, session signals, richer page features), $K$ grows and
the gap between linear (MLP) and quadratic (interaction) savings widens.

\paragraph{Our contributions are as follows:}
\begin{itemize}
  \item \textbf{A rank-aware decomposition principle.} Any linear or
  bilinear operation over a rank-partitioned input admits an exact,
  approximation-free block decomposition. We instantiate this principle
  across the dominant interaction mechanisms in modern ranking
  architectures: FM pairwise interactions (as used in DLRM, DeepFM, NFM, PNN), DCNv2
  cross layers, self-attention, and FC projection layers.

  \item \textbf{Cost asymmetry.} Rank-aware decomposition makes context features
  substantially cheaper to serve than target features. This suggests a new tradeoff between adding context vs. target features. Closed-form analysis reveals this asymmetry, and our empirical ablation shows that increasing $K$ by $3\times$ reduces RPS by $39\%$, while a $3\times$ increase in $M$ reduces it by $64\%$.
  
  \item \textbf{Architecture design principle.} The standard rank-aware decomposition applies only at the first layer in cross networks and attention, because context and target features 
  mix in subsequent layers. This motivates a
  structural design principle: build recommendation architectures from
  layers that admit rank-aware decomposition throughout. We propose rDCN, a
  rank-aware cross network that lifts this restriction across depth, and sketch the analogous
  decomposable attention layer. 

  \item \textbf{Production validation.} Applied without any architectural
  change to a DLRM-style ranker, the decomposition increases per-pod
  throughput by $88\%$ (a $47\%$ reduction in pod count) at identical
  model predictions.
  \end{itemize}

The remainder of the paper is organized as follows.
Section~\ref{sec:related} reviews feature-interaction architectures and
prior work on serving efficiency. Section~\ref{sec:method} presents the
rank-aware decomposition framework and its instantiations for FM, FC,
DCNv2 cross, and attention layers. Section~\ref{sec:rankaware} presents
rank-aware architectural designs: rDCN and a sketch of rank-aware
self-attention. Section~\ref{sec:eval} presents the FLOP analysis,
cross-architecture comparison, controlled ablation, and production
deployment results.
\section{Background and Related Work}
\label{sec:related}

\subsection{Feature Interaction in Recommender Systems}

Modern recommender architectures differ primarily in how they combine features into learned interactions. We briefly review the dominant
families; Section~\ref{sec:method} analyzes each with respect to rank
decomposability.

\paragraph{FM-based neural networks.}
A broad family of architectures uses FM-style pairwise embedding products
as their interaction layer. DeepFM~\cite{guo2017deepfm} combines a
parallel DNN branch with the bilinear pairwise interactions;
NFM~\cite{he2017nfm} introduces bi-interaction pooling followed by an MLP;
PNN~\cite{qu2016pnn} uses inner/outer product layers feeding an MLP.
DLRM~\cite{naumov2019dlrm} is the dominant industrial template: FM-style
pairwise interactions followed by a top MLP.

\paragraph{Cross networks.}
DCN~\cite{wang2017dcn} and DCNv2~\cite{wang2021dcnv2} compute feature crosses
via residual layers of the form
$\x_{l+1} = \x_0 \odot (\W_l \x_l + \bias_l) + \x_l$.
GDCN~\cite{wang2023gdcn} adds information gating. DCNv3~\cite{li2024dcnv3}
extends the formulation with exponential cross networks.

\paragraph{Attention-based interactions.}
AutoInt~\cite{song2019autoint} applies multi-head self-attention over field
embeddings. InterHAt~\cite{li2020interhat} adds hierarchical aggregation.
FiBiNET~\cite{huang2019fibinet} combines SENET-based feature reweighting with
bilinear interactions. MaskNet~\cite{wang2021masknet} introduces instance-guided
masks derived from the full input.

All of the above architectures combine context and target features, creating the
redundancy we address.

\subsection{Making Multi-Target Scoring Lighter}

Prior work attacks this cost from several angles, summarized below.

\paragraph{Structural separation and distillation.}
LightSUAN~\cite{lightsuan2025} explicitly identifies the cost of
multi-target scoring under early-fusion architectures, such as target
attention over user-behavior sequences (DIN \cite{zhou2018din},
ETA \cite{pi2020eta}, TWIN \cite{chang2023twin}) or shared bottom
modules, where context-context computations are repeated $N$ times per
request. It addresses this through distillation: a heavier teacher (SUAN, a
Stacked Unified Attention Network) is compressed into a lighter student
(LightSUAN) that retains early fusion but uses sparse self-attention and
parallel-inference batching, trading capacity for efficiency rather than
restructuring the fusion mechanism.
Two-tower architectures~\cite{yi2019sampling,huang2020embedding} take a
more extreme structural path: independent user and item towers with
scoring reduced to a dot product, efficient enough for retrieval over
millions of candidates but sacrificing joint user--item features and
deep cross-interactions, making them rarely used for final ranking---the
regime our work targets.
Both lines of work trade expressiveness for serving efficiency. Our
rank-aware decomposition contrasts: it is mathematically exact and
preserves expressiveness fully, but applies only where context and
target features remain separately accessible at the interaction layer.
Extending it to early-fused architectures is an open direction for
future work.

\paragraph{Infrastructure-level optimizations.}
ROO~\cite{roo2025} (Meta, 2025) restructures training data so that each
row represents a full user request---one set of user features plus an
array of $N$ candidate items---rather than one $(user, item)$ pair per
row. This deduplicates user features and embedding lookups across the
candidates of a request, yielding 32--100\% training throughput gain for
late-stage ranking.
AIF~\cite{aif2025} executes interaction-independent paths (user-only,
item-only) asynchronously and approximates the skipped cross-interactions
via a learned bridge embedding ($+5.80\%$ RPM at pre-ranking with
negligible added latency). Both operate at the system level and are
complementary to our algebraic approach.

\paragraph{Model-specific interaction-layer reformulations.}
Low Rank FwFM \cite{shtoff2024fwfm} extends the linear-time FM evaluation
trick of Rendle~\cite{rendle2010fm} to the field-weighted variant: it decomposes the Field-Weighted FM (FwFM \cite{pan2018fwfm})
interaction matrix as $R = U^T \text{diag}(e) U + \text{diag}(d)$
(diagonal plus low-rank), and exploits the resulting structure to compute
the context-side sum once per request. Reports $34\%$ average inference
latency reduction in a production advertising system. The approach is
mathematically elegant but scoped to FwFM and requires an approximation
of $R$; it does not extend to FM-based deep networks where pairwise
outputs feed downstream layers.

\paragraph{MaRI.}
Closest to our work, MaRI~\cite{mari2026} (Kuaishou, 2026) automatically
detects redundant FC/MatMul computations via a graph coloring algorithm.
When the input to a MatMul has the tiled form
$\x = [\x_u^{\text{tiled}}; \x_i; \x_c]$, MaRI decomposes the weight matrix
to compute $\text{Tile}(\x_u \W_u) + \x_i \W_i + \x_c \W_c$, reporting
$5.9\%$ hardware savings in Kuaishou production. MaRI's scope is limited
to MatMul(\emph{data}, \emph{weight}) nodes: savings are linear in the
context dimension $O(d_c)$, and the analysis does not cover pairwise FM
interactions (no weight matrix, so no graph node to detect), cross-network
layers, or attention.
\section{Rank-Aware Decomposition}
\label{sec:method}

This section presents our first family of contributions: a systematic,
\emph{identity-equivalent} block decomposition of feature interaction
and projection layers that exploits the rank asymmetry between context
and target features in multi-target scoring. Each decomposition
presented here produces mathematically identical model output to the
standard implementation, just computed more efficiently.
We first state the unifying algebraic principle
(\S\ref{sec:method-unified}), then instantiate it on FM interactions
(\S\ref{sec:method-fm}), FC layers (\S\ref{sec:method-fc}), DCNv2 cross
layers (\S\ref{sec:method-dcn}), and self-attention
(\S\ref{sec:method-attn}), and conclude with a summary of architectural
applicability (\S\ref{sec:method-applicability}).
Section~\ref{sec:rankaware} extends this framework with
\emph{architectural variants}---designs that trade strict identity
equivalence for compounded savings across depth.

\subsection{Problem Setup and Notation}
\label{sec:method-setup}

Consider a recommendation request comprising:
\begin{itemize}
\item \textbf{Context features} $\mathbf{c} = \{c_1, \ldots, c_K\}$ with
embeddings $\e_{c_k} \in \R^D$, identical across all candidates in the request.
We use ``context'' to refer to all features that are constant across
candidates within a request---both user features (profile, history) and
session/page features. Settings in which the same user is scored under multiple
distinct impression contexts in one request can be modeled by treating each
impression as a separate request, or by introducing an additional rank for
the impression axis; we focus on the single-context-per-request case
throughout.
\item \textbf{Candidate items} indexed $n = 1, \ldots, N$, each with target
features $\mathbf{t}^{(n)} = \{t_1^{(n)}, \ldots, t_M^{(n)}\}$ and embeddings
$\e_{t_m}^{(n)} \in \R^D$. When stacked across candidates, we write
$\mathbf{T} \in \R^{N \times M \times D}$ (or $\R^{N \times d_t}$ where
$d_t = M \cdot D$ once embedded and flattened) for the rank-3 tensor of
all $N$ candidates' target features.
\end{itemize}

The model must produce $N$ scores $\hat{y}_1, \ldots, \hat{y}_N$.

\paragraph{Tensor rank convention.} We distinguish features by the rank of
their natural tensor representation:
\begin{itemize}
\item \textbf{Context embeddings:} rank~2, shape $[B, D]$---identical
across candidates.
\item \textbf{Target embeddings:} rank~3, shape $[B, N, D]$---vary per
candidate.
\end{itemize}
Here $B$ is the request batch size (number of requests processed in
parallel). In the remainder of the paper we treat $B = 1$; all FLOP
calculations and savings scale linearly with $B$.

\subsection{The Unifying Principle}
\label{sec:method-unified}

The decompositions in \S\S\ref{sec:method-fm}--\ref{sec:method-attn} are
all instances of a single algebraic principle. Any operation that is
\emph{linear} or \emph{bilinear} in its inputs admits a block decomposition
when the inputs partition as $\x = [\x_c; \x_t]$.

\paragraph{Linear case.} For $f(\x) = \W \x$ with the weight matrix
partitioned by input block, $\W = [\W_c \mid \W_t]$:
\begin{equation}
\W \x = \W_c \x_c + \W_t \x_t
\label{eq:linear-decomp}
\end{equation}
The $\W_c \x_c$ term is computed once when $\x_c$ is rank~2 (shared across
candidates). This single identity
covers FC layers, attention projections, DCNv2's $\W_l \x_l$, and linear
gating mechanisms.

\paragraph{Bilinear case.} For $f(\x, \mathbf{y}) = \x^T \mathbf{W} \mathbf{y}$
with $\x = [\x_c; \x_t]$ and $\mathbf{y} = [\mathbf{y}_c; \mathbf{y}_t]$:
\begin{multline}
\x^T \mathbf{W} \mathbf{y} = \x_c^T \mathbf{W}_{cc} \mathbf{y}_c +
\x_c^T \mathbf{W}_{ct} \mathbf{y}_t + \x_t^T \mathbf{W}_{tc} \mathbf{y}_c +
\x_t^T \mathbf{W}_{tt} \mathbf{y}_t
\label{eq:bilinear-decomp}
\end{multline}
The $\x_c^T \mathbf{W}_{cc} \mathbf{y}_c$ term is computed once. This covers
FM interactions, FwFM, bilinear interactions (FiBiNET~\cite{huang2019fibinet}),
outer products (CIN in xDeepFM~\cite{lian2018xdeepfm}), and attention scores.

\paragraph{Standard approach.} Broadcast (tile) context embeddings to rank~3
before any interaction computation, yielding a unified $[N, (K+M) \cdot D]$
feature tensor. All downstream computation is $O(N)$ regardless of feature type.

\paragraph{Our approach.} Defer broadcasting. Compute context-only operations
at rank~2, target-only and cross operations at rank~3. Broadcast results only
when mixing.

\paragraph{Where the principle fails.}
\begin{enumerate}
\item \textbf{Mixing operations:} When an operation mixes
rank-2 context and rank-3 target inputs into a single output, the output
dimensions are rank-mixed and any downstream layer consuming them sees a
rank-3 ``context'' that is no longer pure context. This is the obstacle
in FC and DCNv2 cross layers beyond layer~0 (\S\ref{sec:method-dcn}).
\item \textbf{Target attention:} DIN~\cite{zhou2018din},
TWIN~\cite{chang2023twin}, ETA~\cite{pi2020eta}---user representations
depend on target by design, so no context-only path exists.
\end{enumerate}

The remaining subsections instantiate this principle on
the dominant feature-interaction mechanisms in modern recommender
architectures; Table~\ref{tab:applicability} (\S\ref{sec:method-applicability})
summarizes coverage across architecture families.

\subsection{FM Interaction Decomposition}
\label{sec:method-fm}

A Factorization Machine~\cite{rendle2010fm} computes pairwise inner products
over all $K + M$ feature embeddings. We follow the modern industrial
formulation in which each $\e_i \in \R^D$ is the field-level embedding for
feature field $i$---obtained by looking up the active categorical value
(or, for multi-valued fields, by pooling the embeddings of the active
values).
This subsumes the classical formulation $\sum_{i<j} \langle v_i, v_j\rangle
x_i x_j$~\cite{rendle2010fm} in which $v_i$ is a latent factor and $x_i$ a
sparse feature value: the field embedding $\e_i$ here plays the role of
$x_i v_i$ for the active value of field $i$, so dense feature values and
one-hot indicators are absorbed into the embedding lookup. In line with
DLRM-style usage, we treat the FM \emph{output} as the vector of pairwise
dot products
\begin{equation}
\text{FM}(\e) = \big[\, \langle \e_i, \e_j \rangle \,\big]_{1 \le i < j \le K+M}
\in \R^{\binom{K+M}{2}}
\label{eq:fm}
\end{equation}
rather than as the scalar sum in the classical formulation; downstream FC
layers consume this vector. The decomposition below operates purely on the
$\e_i$ and is independent of how each embedding is produced.

\paragraph{Rank-aware decomposition.} Applying the bilinear principle
(Eq.~\ref{eq:bilinear-decomp}) to the FM output, we split the pairs into
two groups organized by tensor rank:
\begin{equation}
\text{FM}(\e) =
\big[\,\underbrace{\text{FM}_{\text{ctx}}(\mathbf{c})}_{\text{rank 2, once}}\;;\;
\underbrace{\text{FM}_{\text{tgt}}(\mathbf{c}, \mathbf{T})}_{\text{rank 3, per candidate}}\,\big]
\label{eq:fm-decomp}
\end{equation}
where (concatenation is over pairs)
\begin{align}
\text{FM}_{\text{ctx}}(\mathbf{c}) &=
  \big[\, \langle \e_{c_i}, \e_{c_j} \rangle \,\big]_{1 \le i<j \le K}
  \in \R^{\binom{K}{2}} \label{eq:fm-ctx} \\
\text{FM}_{\text{tgt}}(\mathbf{c}, \mathbf{T}^{(n)}) &=
  \big[\, \langle \e_{t_i}^{(n)}, \e_{c_k} \rangle \,\big]_{i,k}
  \,\Big|\,
  \big[\, \langle \e_{t_i}^{(n)}, \e_{t_j}^{(n)} \rangle \,\big]_{i<j}
  \label{eq:fm-tgt}
\end{align}

$\text{FM}_{\text{ctx}}$ depends only on context embeddings and is
identical for every candidate---computed once per request.
$\text{FM}_{\text{tgt}}$ captures every interaction involving \emph{at
least one} target field: both target--target pairs and context--target pairs.
We group these together because they share the same rank (3), the same
per-candidate cost structure, and---as we will show---admit a single
asymmetric matmul implementation. This ``target-anchored'' grouping is
structurally the same pattern we will see again in rDCN
(\S\ref{sec:rdcn}) and in rank-aware attention
(\S\ref{sec:rankaware-attn}).

\begin{table}[t]
\centering
\caption{Rank-aware FM decomposition.}
\label{tab:fm-groups}
\begin{tabular}{llll}
\toprule
Component & Count & Rank & Per request \\
\midrule
$\text{FM}_{\text{ctx}}$ & $K(K-1)/2$ & 2 & Once \\
$\text{FM}_{\text{tgt}}$ & $K M + M(M-1)/2$ & 3 & $N$ times \\
\bottomrule
\end{tabular}
\end{table}

\paragraph{Implementation.} Two building blocks realize this decomposition:
\begin{itemize}
\item $\text{FM}_{\text{ctx}}$ computes as the
upper-triangle of the $K \times K$ context inner-product matrix. All
operands are rank 2; the result is a rank-2 tensor of size $\binom{K}{2}$
computed once per request.
\item $\text{FM}_{\text{tgt}}$ computes
in a single asymmetric matmul: each target field embedding is multiplied
against the stacked $[\mathbf{c}; \mathbf{T}^{(n)}]$ block, covering both
the context--target and target--target pairs simultaneously. Context
embeddings are broadcast to rank~3 only within this operator.
\end{itemize}

\paragraph{FLOP savings.} The standard implementation computes all
$\binom{K+M}{2}$ interactions at rank 3, costing
$N \cdot \binom{K+M}{2} \cdot D$ FLOPs per request. Our decomposition
costs $\binom{K}{2} \cdot D + N \cdot \big( K M + \binom{M}{2} \big) \cdot D$.
The absolute savings are:
\begin{equation}
\Delta_{\text{FM}} = (N-1) \cdot \binom{K}{2} \cdot D
\label{eq:fm-savings}
\end{equation}
a quadratic function of $K$. The savings fraction of total FM FLOPs is:
\begin{equation}
\frac{\Delta_{\text{FM}}}{\text{FLOPs}^{\text{std}}_{\text{FM}}} =
\frac{(N-1) \cdot K(K-1)}{N \cdot (K+M)(K+M-1)}
\label{eq:fm-savings-fraction}
\end{equation}
The savings fraction is larger when $M$ is smaller.

\paragraph{Applicability across FM variants.} The decomposition above is
written for the standard FM formulation but applies unchanged to any
architecture whose interaction layer computes pairwise embedding products
between feature fields, including DLRM~\cite{naumov2019dlrm},
DeepFM~\cite{guo2017deepfm}, NFM~\cite{he2017nfm}, and PNN~\cite{qu2016pnn}.
For field-aware variants such as FFM~\cite{juan2016ffm}, the same two-way
decomposition holds with field-specific embeddings substituted on each side
of the inner product.

\subsection{FC Layer Decomposition}
\label{sec:method-fc}

Recommender models commonly stack one or more FC layers after the
interaction layer. We treat the first FC layer's input as a concatenation
of a rank-2 context part $\x_c$ and a rank-3 target part $\x_t$---either
passed directly or, as in DLRM-style models, produced as the outputs of
$\text{FM}_{\text{ctx}}$ and $\text{FM}_{\text{tgt}}$. The rank-aware FC
decomposition is a direct application of the linear case of
\S\ref{sec:method-unified} (Eq.~\ref{eq:linear-decomp}). The first FC
layer exploits the block structure:
\begin{align}
\mathbf{h} &= \sigma\big([\x_c; \x_t] \W + \bias\big) \\
&= \sigma\big(\x_c \W_c + \x_t \W_t + \bias\big)
\end{align}
where $\x_c$ is rank-2, $\x_t$ is rank-3, and
$\W = [\W_c; \W_t]$ is the kernel partitioned along the input
dimension. It is the same $\W_c, \W_t$ split as in
Equation~\ref{eq:linear-decomp} (here we write the matrix as concatenated
row blocks because the input is concatenated; the algebra is identical).

\paragraph{RankSplitDense.}
We compute $\x_c \W_c$ once at rank~2, broadcast to rank~3, and add to
the per-candidate $\x_t \W_t$ before applying the nonlinearity. Savings:
\begin{equation}
\Delta_{\text{FC}} = (N-1) \cdot |\x_c| \cdot U
\label{eq:fc-savings}
\end{equation}
where $U$ is the number of units in the FC layer and $|\x_c|$ is the size
of the rank-2 context input.

\paragraph{Why only the first FC layer.} After
$\sigma(\x_c\W_c + \x_t\W_t + \bias)$, the context and target
signals are mixed at every neuron by the nonlinear activation. All
subsequent layers operate on fully target-dependent tensors of shape
$[N, U]$; no further rank-separated computation is possible without
approximation.

When the FC layer sits
\emph{downstream of a rank-aware FM}, the context-only FM output
$\x_c = \text{FM}_{\text{ctx}}(\mathbf{c})$ has size $\binom{K}{2}$
(Eq.~\ref{eq:fm-ctx})---itself quadratic in the context field count.
Substituting into Equation~\ref{eq:fc-savings}:
\begin{equation}
\Delta_{\text{FC}} = (N-1) \cdot \binom{K}{2} \cdot U
\label{eq:fc-savings-compounded}
\end{equation}
also scales \emph{quadratically} with $K$.

Combined with the FM-layer savings of Equation~\ref{eq:fm-savings}, a
DLRM-style model thus enjoys quadratic-in-$K$ savings at \emph{both} stages
of the forward pass: once in the pairwise interaction layer where the
$\binom{K}{2}$ context-context pairs are computed, and again in the first
FC layer where those pairs are projected. This compounding is what makes
the optimization so impactful for interaction-heavy models and is absent in
MLP-only architectures, where the context input to FC scales only
linearly with $K$.

\subsection{DCNv2 Cross Layer Decomposition}
\label{sec:method-dcn}

A DCNv2 cross layer~\cite{wang2021dcnv2} has the form:
\begin{equation}
\x_{l+1} = \x_0 \odot (\W_l \x_l + \bias_l) + \x_l
\label{eq:dcnv2}
\end{equation}
The matmul $\W_l \x_l$ is precisely the block-output linear case of
\S\ref{sec:method-unified} (Eq.~\ref{eq:linear-decomp}), with the input
partition $\x_l = [\x_l^c; \x_l^t]$ and the weight matrix partitioned
into four blocks:
\begin{equation}
\W_l \x_l =
\begin{bmatrix} \W_{cc} & \W_{ct} \\ \W_{tc} & \W_{tt} \end{bmatrix}
\begin{bmatrix} \x_l^c \\ \x_l^t \end{bmatrix} =
\begin{bmatrix} \W_{cc} \x_l^c + \W_{ct} \x_l^t \\ \W_{tc} \x_l^c + \W_{tt} \x_l^t \end{bmatrix}
\end{equation}
At layer $l=0$ (where $\x_l = \x_0 = [\x_0^c; \x_0^t]$), the terms
$\W_{cc} \x_0^c$ and $\W_{tc} \x_0^c$ depend only on context features and are computable once per
request. Savings fraction for the first cross layer:
\begin{equation}
\frac{d_c^2}{(d_c + d_t)^2}
\end{equation}

\paragraph{Why depth matters: the cross-block matmul re-entangles ranks.}
The savings above apply only at layer~0. The blocker beyond layer~0 is the
same one identified as the second failure mode in
\S\ref{sec:method-unified}: any linear operation whose inputs include both
rank-2 context and rank-3 target produces output rows whose ``context
dimensions'' depend on target features. Concretely, the context-output rows
of $\W_l \x_l$ are $\W_{cc} \x_l^c + \W_{ct} \x_l^t$. Once layer~0 emits this rank-mixed output, every
subsequent layer's $\x_l^c$ is target-dependent, and $\W_{cc} \x_l^c$ is no
longer a rank-2 quantity that can be precomputed. The Hadamard product
$\x_0 \odot (\cdot)$ that follows the matmul is incidental: it preserves
rank-3-ness but does not create it. Replacing $\odot$ with a sum or any
other elementwise operator would not recover rank separation, because the
rank mixing has already happened in $\W_l \x_l$ itself. Beyond layer~0,
practical savings diminish to zero. Section~\ref{sec:rdcn}
proposes rDCN, a redesigned cross layer that prevents this re-entanglement.

\paragraph{Compatibility with DCNv2's low-rank parameterization.}
The original DCNv2 paper~\cite{wang2021dcnv2} also proposes a low-rank
parameterization $\W_l = U_l V_l^\top$ with $U_l, V_l \in \R^{D \times r}$
to reduce parameter count. This composes directly with the rank-aware
block decomposition: partitioning the rows as $U_l = [U_c; U_t]$ and
$V_l = [V_c; V_t]$ induces the block factorization
$\W_{cc} = U_c V_c^\top$ (and analogously for $\W_{ct}, \W_{tc}, \W_{tt}$)
at no extra parameter cost, and the once-per-request context matmul
$\W_{cc}\x_0^c = U_c(V_c^\top \x_0^c)$ remains rank~2.

\subsection{Attention Layer Decomposition}
\label{sec:method-attn}

Attention-based interaction layers (e.g., AutoInt~\cite{song2019autoint},
InterHAt~\cite{li2020interhat}) apply self-attention over field embeddings:
\begin{equation}
\text{Attn}(Q, K, V) = \text{softmax}\!\left(\frac{QK^T}{\sqrt{d_k}}\right)\!V
\end{equation}
with $Q = \W_Q E$, $K = \W_K E$, $V = \W_V E$, where $E$ is the stacked field
embedding matrix.

\paragraph{Q/K/V projections decompose directly.} Each projection is a MatMul
over concatenated field embeddings and is an instance of
Equation~\ref{eq:linear-decomp}:
\begin{equation}
\W_Q E = \W_Q [E_c; E_t] = \W_{Q,c} E_c + \W_{Q,t} E_t
\end{equation}
The context-side projections $\W_{Q,c} E_c$, $\W_{K,c} E_c$, $\W_{V,c} E_c$
are computed once per request. Target-side projections are per candidate.

\paragraph{Attention matrix has block structure.} With fields partitioned into
context ($C$) and target ($T$), the full attention matrix decomposes via the
bilinear case (Eq.~\ref{eq:bilinear-decomp}):
\begin{equation}
A = \text{softmax}\!\left(
\begin{bmatrix} Q_c K_c^T & Q_c K_t^T \\ Q_t K_c^T & Q_t K_t^T \end{bmatrix}\right)
\end{equation}
The $Q_c K_c^T$ block (context queries attending to context keys) is rank~2
and computed once. The remaining blocks are rank~3.

\paragraph{Softmax caveat.} The softmax normalization mixes context
and target contributions in each row's denominator. However, the
unnormalized context-context block is precomputable, and the row
softmax can be assembled via FlashAttention-style block-wise
log-sum-exp---combining precomputed context partial sums with
per-candidate target contributions in $O(1)$ extra work per query
position.

Savings in attention scale with the fraction of operations involving only
context fields, approximately $K^2 / (K+M)^2$ for the score computation,
again quadratic in $K$. Due to the final mixing of different ranks in the attention layer, these
savings, again, apply only to the first layer.

\subsection{Applicability Across Architectures}
\label{sec:method-applicability}

Table~\ref{tab:applicability} summarizes which architecture families admit
rank-aware decomposition. The decomposition applies fully to DLRM-style
architectures (the dominant industrial paradigm), partially to
cross-network-based architectures (limited beyond layer~0 by the
cross-block matmul; see \S\ref{sec:rdcn} for the rDCN redesign that lifts
this restriction), and to attention-based architectures (with the softmax
caveat).

\begin{table}[t]
\centering
\caption{Applicability of rank-aware decomposition to the interaction
layer across popular recommender architecture families. The FC stack
admits decomposition in all cases and is omitted.}
\label{tab:applicability}
\small
\begin{tabular}{ll}
\toprule
Architecture & Interaction layer \\
\midrule
DLRM~\cite{naumov2019dlrm} / DeepFM~\cite{guo2017deepfm}     & FM pairs: Full \\
FM~\cite{rendle2010fm} / FFM~\cite{juan2016ffm} / NFM~\cite{he2017nfm} / PNN~\cite{qu2016pnn}
                                                            & FM/bilinear: Full \\
FinalMLP~\cite{zhu2023finalmlp}                             & Two-stream: Full \\
\midrule
DCNv2~\cite{wang2021dcnv2} / GDCN~\cite{wang2023gdcn}       & Cross: first layer only \\
\midrule
AutoInt~\cite{song2019autoint} / InterHAt~\cite{li2020interhat}
                                                            & Attention: partial \\
FiBiNET~\cite{huang2019fibinet}                             & Bilinear: partial \\
MaskNet~\cite{wang2021masknet}                              & Gated: partial \\
xDeepFM~\cite{lian2018xdeepfm}                              & Outer products: partial \\
DHEN~\cite{zhang2022dhen}                                   & Per-submodule \\
\bottomrule
\end{tabular}
\end{table}

\section{Rank-Aware Architecture Design}
\label{sec:rankaware}

The rank-aware, identity-equivalent decompositions of \S\ref{sec:method}
optimize existing architectures without changing their predictions, but
their savings are bounded: in cross networks and attention, each layer
mixes ranks in its output, leaving subsequent layers without
rank-separated inputs to decompose, so the decomposition applies only
at the first layer. This section introduces a complementary class of
\emph{architectural variants} that lift this restriction by accepting a
constraint on the layer form to avoid rank-mixing. The variants are
not identity-equivalent to their standard analogues, but offline
evaluation shows accuracy within training noise alongside savings that
compound across depth (\S\ref{sec:eval-comparison}).

\subsection{The Asymmetric Loading Principle}
\label{sec:rankaware-principle}

The rank-mixing has a specific source in each architecture: in DCNv2,
the off-diagonal $\W_{ct}$ block maps rank-3 target inputs into the
context-dimension half of the layer state; in self-attention,
context-field outputs attend over target-field keys and values, again
producing context outputs that depend on every candidate.

This motivates the \emph{asymmetric loading principle}: architectures
should load context$\times$target interactions onto the target stream
only, while the context stream evolves purely through
context$\times$context dynamics---so rank separation propagates across
the full depth of the stack. Whether this directional constraint
meaningfully limits expressiveness is an empirical question that requires
hyperparameter tuning of both architectures to compare properly. We instantiate this principle for cross networks
(\S\ref{sec:rdcn}) and attention (\S\ref{sec:rankaware-attn}), with
initial offline results for the cross-network instantiation reported in
\S\ref{sec:eval-comparison}.

The principle composes with the FC decomposition of
\S\ref{sec:method-fc}: any rank-aware interaction stack---FM
(\S\ref{sec:method-fm}) or the architectural variants introduced
below---terminates in a rank-2 context representation and a rank-3
target representation, which feed directly into a rank-decomposed FC layer.

\subsection{Rank-Aware DCN: rDCN}
\label{sec:rdcn}

We instantiate the asymmetric loading principle for DCNv2 cross networks
by structurally removing the $\W_{ct}$ block (\S\ref{sec:method-dcn}) and
maintaining two separately-typed states---a rank-2 context stream
$\mathbf{c}_l$ and a rank-3 target stream $\mathbf{T}_l$---each updated
in parallel by its own DCNv2-style layer:
\begin{align}
\mathbf{T}_{l+1} &= \mathbf{T}_0 \odot \Big(
  \underbrace{\W^{\mathrm{ct}}_l\, \mathbf{c}_l}_{\text{rank 2, once}}
  \;\oplus\;
  \underbrace{\W^{\mathrm{t}}_l\, \mathbf{T}_l}_{\text{rank 3}}
  + \bias^{\mathrm{t}}_l \Big) + \mathbf{T}_l
  \label{eq:rdcn-t} \\
  \mathbf{c}_{l+1} &= \mathbf{c}_0 \odot
  (\W^{\mathrm{c}}_l\, \mathbf{c}_l + \bias^{\mathrm{c}}_l) + \mathbf{c}_l
  \label{eq:rdcn-c}
\end{align}
with $\W^{\mathrm{c}}_l \in \R^{D_c \times D_c}$,
$\W^{\mathrm{ct}}_l \in \R^{D_t \times D_c}$,
$\W^{\mathrm{t}}_l \in \R^{D_t \times D_t}$, and $\oplus$ broadcast-and-add
along the $N$ axis. Equivalently, for each candidate $j$:
$\mathbf{T}_{l+1}^{(j)} = \mathbf{T}_0^{(j)} \odot
(\W^{\mathrm{ct}}_l \mathbf{c}_l + \W^{\mathrm{t}}_l \mathbf{T}_l^{(j)} + \bias^{\mathrm{t}}_l) + \mathbf{T}_l^{(j)}$.

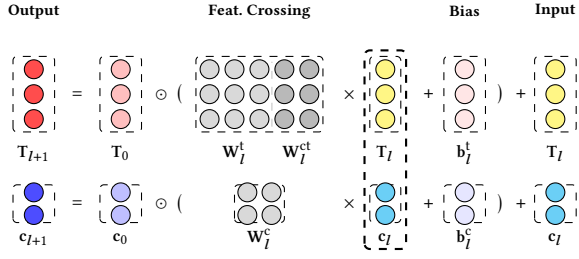
\begin{figure}[t]
\centering
\begin{tikzpicture}[
  font=\scriptsize,
  Tout/.style={circle, draw, fill=red!70,    minimum size=2.5mm, inner sep=0pt},
  T0c/.style={circle, draw, fill=red!25,    minimum size=2.5mm, inner sep=0pt},
  Tlc/.style={circle, draw, fill=yellow!60, minimum size=2.5mm, inner sep=0pt},
  coutc/.style={circle, draw, fill=blue!70, minimum size=2.5mm, inner sep=0pt},
  c0c/.style={circle, draw, fill=blue!25,   minimum size=2.5mm, inner sep=0pt},
  clc/.style={circle, draw, fill=cyan!50,   minimum size=2.5mm, inner sep=0pt},
  Wcell/.style={circle, draw, fill=gray!30, minimum size=2.5mm, inner sep=0pt},
  Wctcell/.style={circle, draw, fill=gray!55, minimum size=2.5mm, inner sep=0pt},
  bTc/.style={circle, draw, fill=red!10,    minimum size=2.5mm, inner sep=0pt},
  bcc/.style={circle, draw, fill=blue!10,   minimum size=2.5mm, inner sep=0pt},
  tbox/.style={draw, dashed, rounded corners=1.5pt},
  tboxouter/.style={draw, dashed, rounded corners=2pt, thick},
  hd/.style={font=\bfseries\scriptsize},
]

% ===== HEADERS =====
\node[hd] at (0.475, 1.10) {Output};
\node[hd] at (3.46,  1.10) {Feat.\ Crossing};
\node[hd] at (6.175, 1.10) {Bias};
\node[hd] at (7.375, 1.10) {Input};

% ===== TOP ROW (target stream, y = 0) =====

% T_{l+1}
\draw[tbox] (0.20, -0.50) rectangle (0.75, 0.50);
\foreach \y in {-0.33, 0, 0.33} \node[Tout] at (0.475, \y) {};
\node at (0.475, -0.78) {$\mathbf{T}_{l+1}$};

\node at (1.05, 0) {$=$};

% T_0
\draw[tbox] (1.35, -0.50) rectangle (1.90, 0.50);
\foreach \y in {-0.33, 0, 0.33} \node[T0c] at (1.625, \y) {};
\node at (1.625, -0.78) {$\mathbf{T}_0$};

\node at (2.16, 0) {$\odot$};
\node at (2.40, 0) {$\big($};

% [W^t | W^{ct}] block, 3 rows x 5 cols
\draw[tbox] (2.62, -0.50) rectangle (4.30, 0.50);
\foreach \x in {2.80, 3.13, 3.46} \foreach \y in {-0.33, 0, 0.33}
  \node[Wcell] at (\x, \y) {};
\foreach \x in {3.79, 4.12} \foreach \y in {-0.33, 0, 0.33}
  \node[Wctcell] at (\x, \y) {};
\draw[dashed, gray!50, very thin] (3.625, -0.43) -- (3.625, 0.43);
\node at (3.13, -0.78) {$\W^{\mathrm{t}}_l$};
\node at (3.96, -0.78) {$\W^{\mathrm{ct}}_l$};

\node at (4.65, 0) {$\times$};

% ===== TALL CONCAT BOX =====
\draw[tboxouter] (4.85, -2.05) rectangle (5.40, 0.58);
% Inner T_l frame (top-row level)
\draw[tbox] (4.92, -0.50) rectangle (5.33, 0.50);
\foreach \y in {-0.33, 0, 0.33} \node[Tlc] at (5.125, \y) {};
\node at (5.125, -0.78) {$\mathbf{T}_l$};
% Inner c_l frame (bottom-row level)
\draw[tbox] (4.92, -1.70) rectangle (5.33, -1.20);
\foreach \y in {-1.60, -1.30} \node[clc] at (5.125, \y) {};
\node at (5.125, -1.92) {$\mathbf{c}_l$};

\node at (5.70, 0) {$+$};

% b^t
\draw[tbox] (5.90, -0.50) rectangle (6.45, 0.50);
\foreach \y in {-0.33, 0, 0.33} \node[bTc] at (6.175, \y) {};
\node at (6.175, -0.78) {$\bias^{\mathrm{t}}_l$};

\node at (6.65, 0) {$\big)$};
\node at (6.92, 0) {$+$};

% T_l residual
\draw[tbox] (7.10, -0.50) rectangle (7.65, 0.50);
\foreach \y in {-0.33, 0, 0.33} \node[Tlc] at (7.375, \y) {};
\node at (7.375, -0.78) {$\mathbf{T}_l$};

% ===== BOTTOM ROW (context stream, y = -1.45) =====

% c_{l+1}
\draw[tbox] (0.20, -1.70) rectangle (0.75, -1.20);
\foreach \y in {-1.60, -1.30} \node[coutc] at (0.475, \y) {};
\node at (0.475, -1.92) {$\mathbf{c}_{l+1}$};

\node at (1.05, -1.45) {$=$};

% c_0
\draw[tbox] (1.35, -1.70) rectangle (1.90, -1.20);
\foreach \y in {-1.60, -1.30} \node[c0c] at (1.625, \y) {};
\node at (1.625, -1.92) {$\mathbf{c}_0$};

\node at (2.16, -1.45) {$\odot$};
\node at (2.40, -1.45) {$\big($};

% W^c (2x2), centered at x=3.46 (under Feat. Crossing)
\draw[tbox] (3.13, -1.70) rectangle (3.79, -1.20);
\foreach \x in {3.30, 3.62} \foreach \y in {-1.60, -1.30}
  \node[Wcell] at (\x, \y) {};
\node at (3.46, -1.92) {$\W^{\mathrm{c}}_l$};

% x symbol aligned with top row's x (at x=4.65)
\node at (4.65, -1.45) {$\times$};

% (matmul input is the c_l circles in the concat box at x=5.125, y=-1.45)

\node at (5.70, -1.45) {$+$};

% b^c
\draw[tbox] (5.90, -1.70) rectangle (6.45, -1.20);
\foreach \y in {-1.60, -1.30} \node[bcc] at (6.175, \y) {};
\node at (6.175, -1.92) {$\bias^{\mathrm{c}}_l$};

\node at (6.65, -1.45) {$\big)$};
\node at (6.92, -1.45) {$+$};

% c_l residual (same shared c_l, redrawn as tensor box)
\draw[tbox] (7.10, -1.70) rectangle (7.65, -1.20);
\foreach \y in {-1.60, -1.30} \node[clc] at (7.375, \y) {};
\node at (7.375, -1.92) {$\mathbf{c}_l$};

\end{tikzpicture}
\caption{One rDCN layer. Top row: target stream. Bottom row: context
stream. The $\mathbf{c}_l$ tensor is shared between the two streams.}
\label{fig:rdcn-layer}
\end{figure}

Figure~\ref{fig:rdcn-layer} visualizes one rDCN layer using the
equivalent block-matrix form for the target stream:
$[\W^{\mathrm{t}}_l\,|\,\W^{\mathrm{ct}}_l]\,[\mathbf{T}_l;\mathbf{c}_l]$,
where the $\mathbf{c}_l$ slice of the concatenated input also serves as
the input to the context stream's cross layer below.

The context stream (Eq.~\ref{eq:rdcn-c}) is a standard rank-2 DCNv2 stack
on $\mathbf{c}_l$ alone, evolving the context representation through
higher-order ctx$\times$ctx crosses across depth. The target stream
(Eq.~\ref{eq:rdcn-t}) is the asymmetric layer: it reads the evolved
context $\mathbf{c}_l$ via the rank-2 cross matmul
$\W^{\mathrm{ct}}_l \mathbf{c}_l$ (computed once per request and
broadcast along $N$), combines it with the per-candidate state
$\W^{\mathrm{t}}_l \mathbf{T}_l$, and applies the Hadamard with
$\mathbf{T}_0$. Since no weight matrix produces context-dimension output
from target inputs, the two streams stay rank-2 and rank-3 respectively
at every depth. Empirically, this parallel context stream proves
essential: removing it causes non-trivial accuracy degradation
(Table~\ref{tab:comparison}, last DCN row) despite negligible FLOP
savings, confirming
that the higher-order ctx$\times$ctx dynamics it captures are
load-bearing.

\paragraph{Parameters per layer.}
In the four-block partition of \S\ref{sec:method-dcn}, rDCN structurally
omits the $\W_{ct}$ block (target inputs producing context-dimension
output): the context stream reads only $\mathbf{c}_l$, so no
target-to-context coupling is needed. Baseline DCNv2 has $(D_c + D_t)^2$
weight parameters per layer; rDCN's three matrices total
$D_c^2 + D_c D_t + D_t^2$, a saving of $D_c D_t$ per layer.

\paragraph{FLOPs per layer.} The context stream contributes $\Theta(D_c^2)$
(independent of $N$). The target stream contributes
$\Theta(D_t D_c)$ for the once-per-request cross matmul plus
$\Theta(N D_t^2)$ for the per-candidate target matmul. Baseline DCNv2's
per-layer cost, without first-layer rank-aware optimization, is
$\Theta(N D^2) = \Theta(N (D_c^2 + 2 D_c D_t + D_t^2))$, where
$D = D_c + D_t$. In rDCN this collapses to
$\Theta(D_c^2) + \Theta(N (D_t D_c + D_t^2))$: the context-driven
matmul is removed from the $N$-loop entirely, and the remaining
per-candidate work scales linearly in the context dimension where the
baseline scales quadratically.

\subsection{Rank-Aware Attention}
\label{sec:rankaware-attn}

The same two-stream design applies to attention-based interaction stacks.
As in rDCN, we maintain a rank-2 context stream $\mathbf{c}_l$ and a
rank-3 target stream $\mathbf{T}_l$, each updated by its own attention
sub-layer.

\paragraph{Context stream.}
A standard rank-2 self-attention layer over context field embeddings:
queries, keys, and values are all drawn from $\mathbf{c}_l$. Cost per
layer is $\Theta(D_c^2)$, independent of $N$.

\paragraph{Target-anchored stream.}
Queries are drawn only from the target side,
$\mathbf{Q} = \mathbf{T}_l \W_Q$ (rank 3). Keys and values are split by
source: $\mathbf{K}^c = \mathbf{c}_l \W_K^c$ and
$\mathbf{V}^c = \mathbf{c}_l \W_V^c$ at rank 2 (computed once per
request), alongside $\mathbf{K}^t = \mathbf{T}_l \W_K^t$ and
$\mathbf{V}^t = \mathbf{T}_l \W_V^t$ at rank 3. Target queries attend to
the concatenation $[\mathbf{K}^c; \mathbf{K}^t]$, and the output is a
weighted sum of $[\mathbf{V}^c; \mathbf{V}^t]$. Because context keys and
values are produced at rank 2, their projection cost is amortized across
all $N$ candidates.

A production implementation and empirical
evaluation are left to future work.
\section{Evaluation}
\label{sec:eval}

We evaluate the rank-aware decomposition in three stages. First, we
compare rank-aware component contributions across two architecture
families (DCN-style and DLRM-style) on the production dataset, reporting
FLOPs and offline metrics (\S\ref{sec:eval-comparison}). Second, we characterize per-pod
throughput on synthetic models by sweeping feature counts on production
hardware (\S\ref{sec:eval-ablation}). Third, we report production
deployment results from Taboola's recommendation platform
(\S\ref{sec:eval-production}).

\subsection{Component Contribution Comparison}
\label{sec:eval-comparison}

Table~\ref{tab:comparison} compares identity-equivalent rank-aware
decomposition (\S\ref{sec:method}) and the architectural variant rDCN
(\S\ref{sec:rdcn}) across two architecture families on our production
dataset. Within each family, all configurations share the same model
shape; the families themselves operate in different regimes and are
listed separately.

\begin{table}[!t]
\centering
\caption{Component contributions across architecture families.
$\Delta$ FLOPs are percent change relative to each family's vanilla
baseline. LogLoss is the absolute value with the parenthesized percent
change from vanilla. Variant names list the components with rank-aware
optimization applied (``+FC'' indicates rank-aware FC on top of the
interaction-layer variant); rDCN** denotes rDCN with the parallel
context stream removed.}
\label{tab:comparison}
\small
\begin{tabular*}{\columnwidth}{@{\extracolsep{\fill}}lrrrr@{}}
\toprule
& \multicolumn{3}{c}{$\Delta$ FLOPs} & \\
\cmidrule(lr){2-4}
Variant & Int. & FC & Total & LogLoss ($\Delta\%$) \\
\midrule
\multicolumn{5}{l}{\textit{DCN-style ($K{=}26$, $M{=}16$, $d_c{=}514$, $d_t{=}577$, $L{=}4$)}} \\
DCNv2 + FC      & $-18\%$ & $-36\%$ & $-20\%$ & $0.3876$ ($+0.02\%$) \\
rDCN + FC       & $-72\%$ & $-36\%$ & $-67\%$ & $0.3877$ ($+0.05\%$) \\
rDCN** + FC     & $-72\%$ & $-36\%$ & $-67\%$ & $0.3886$ ($+0.27\%$) \\
\midrule
\multicolumn{5}{l}{\textit{DLRM-style ($K{=}27$, $M{=}4$, $d_c{=}3456$, $d_t{=}512$)}} \\
FM              & $-87\%$ & $0\%$   & $-17\%$ & $0.3875$ ($+0.03\%$) \\
FM + FC         & $-87\%$ & $-69\%$ & $-73\%$ & $0.3875$ ($+0.01\%$) \\
\bottomrule
\end{tabular*}
\end{table}

Within each family, the identity-equivalent rank-aware decomposition reproduces
vanilla predictions to within noise levels ($\pm 0.0001$ logloss),
as expected from a mathematically exact rewriting.
The architectural variant rDCN+FC trades strict identity equivalence for a
much smaller compute and parameter footprint: it matches DCNv2 within
$\pm 0.0002$ logloss while using $67\%$ fewer total FLOPs and without
the $\W_{ct}$ block. The parallel context stream is essential: removing
it (last DCN row) yields negligible additional FLOP savings but
degrades logloss by $\sim 0.001$---an order of magnitude larger than
the within-noise differences elsewhere in the table, confirming that
the $c{\times}c$ interactions the parallel stream captures are
load-bearing.

The throughput and production evaluations that follow focus on the
DLRM-style architecture currently in production; rDCN has been validated
offline above but is not yet deployed online. While rDCN substantially
reduces compute within the DCN family (Table~\ref{tab:comparison}), the
DLRM-style FM+FC variant reaches the same offline accuracy at lower
total compute. Looking forward, rDCN's $\Theta(D_t D_c)$ per-layer cost makes
expanding context within the DCN family (currently $d_c{=}514$ vs.\
DLRM-style's $3456$) substantially cheaper than DCNv2 admits---an
exploration we leave to future work.

\subsection{Throughput Characterization}
\label{sec:eval-ablation}

We validate the theory empirically on synthetic DLRM-style models (FM
interaction layer followed by FC layers), varying one feature-count
parameter at a time. Each configuration is built twice---once with the
vanilla implementation and once with the rank-aware decomposition---and
both are served on the same hardware: a 16-vCPU pod with 16~GiB RAM,
pinned to an Intel Xeon Silver 4510 host (Sapphire Rapids,
2.4--4.1~GHz, AMX/AVX-512 VNNI), running TensorFlow Serving 2.x with
oneDNN enabled. The client issues 64 concurrent in-flight requests in a closed loop
(a new request is dispatched whenever a response returns), and we record
the steady-state request rate (RPS) per configuration. We report \emph{normalized RPS}: each measurement is divided
by the RPS of the smallest-model vanilla baseline,
$(K=8, M=4)$, so that the baseline reads $1.00$ and all other curves can
be read relative to it.

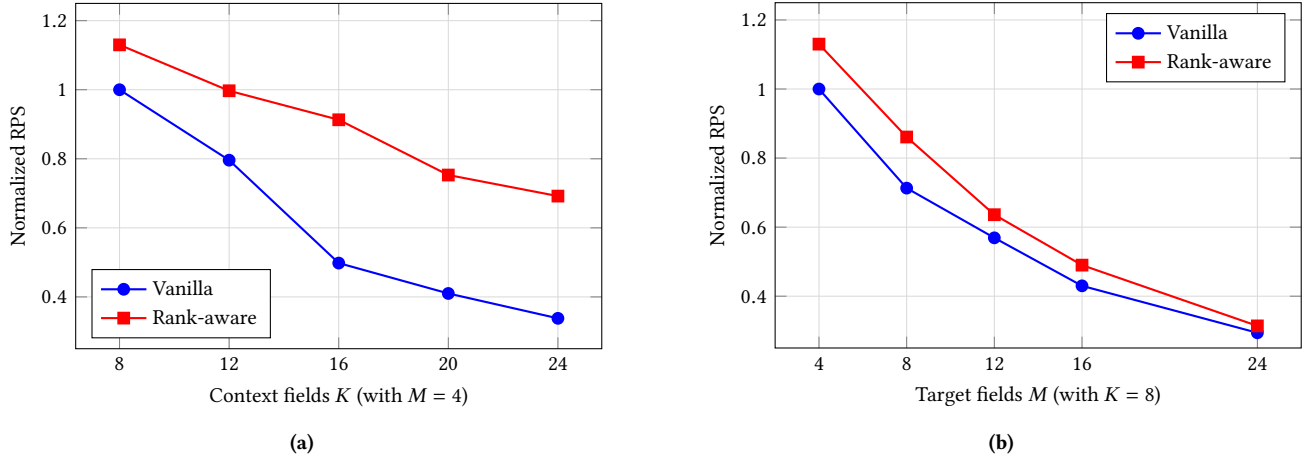
\begin{figure*}[!t]
\centering
\begin{subfigure}[t]{0.48\textwidth}
\centering
\begin{tikzpicture}
\begin{axis}[
  width=\textwidth, height=0.72\textwidth,
  xlabel={Context fields $K$ (with $M=4$)},
  ylabel={Normalized RPS},
  xtick={8,12,16,20,24},
  ymin=0.25, ymax=1.25,
  grid=both, grid style={line width=0.2pt, draw=gray!30},
  legend pos=south west,
  legend cell align={left},
  legend style={font=\small},
  tick label style={font=\small},
  label style={font=\small},
]
\addplot[mark=*, blue, thick]
  coordinates {(8,1.000) (12,0.796) (16,0.498) (20,0.410) (24,0.338)};
\addplot[mark=square*, red, thick]
  coordinates {(8,1.130) (12,0.997) (16,0.913) (20,0.753) (24,0.692)};
\legend{Vanilla, Rank-aware}
\end{axis}
\end{tikzpicture}
\caption{}
\label{fig:ablation-K}
\end{subfigure}
\hfill
\begin{subfigure}[t]{0.48\textwidth}
\centering
\begin{tikzpicture}
\begin{axis}[
  width=\textwidth, height=0.72\textwidth,
  xlabel={Target fields $M$ (with $K=8$)},
  ylabel={Normalized RPS},
  xtick={4,8,12,16,24},
  ymin=0.25, ymax=1.25,
  grid=both, grid style={line width=0.2pt, draw=gray!30},
  legend pos=north east,
  legend cell align={left},
  legend style={font=\small},
  tick label style={font=\small},
  label style={font=\small},
]
\addplot[mark=*, blue, thick]
  coordinates {(4,1.000) (8,0.713) (12,0.569) (16,0.430) (24,0.294)};
\addplot[mark=square*, red, thick]
  coordinates {(4,1.130) (8,0.861) (12,0.636) (16,0.490) (24,0.314)};
\legend{Vanilla, Rank-aware}
\end{axis}
\end{tikzpicture}
\caption{}
\label{fig:ablation-M}
\end{subfigure}
\caption{Throughput normalized to the $(K=8, M=4)$ vanilla baseline.
Left panel (a) varies $K$ with $M{=}4$ fixed; right panel (b) varies
$M$ with $K{=}8$ fixed. The vertical gap between the two curves in each
panel is the throughput gain delivered by rank-aware decomposition at
that configuration.}
\label{fig:ablation}
\end{figure*}

\paragraph{Experiment A: Varying context features $K$ (with $M=4$ fixed).}
Figure~\ref{fig:ablation-K} plots normalized RPS as $K$ grows from 8 to
24. The vanilla curve degrades steeply (from 1.00 to 0.34), while the
rank-aware curve degrades much more gradually (from 1.13 to 0.69), so
the rank-aware advantage \emph{widens} from $+13\%$ at $K=8$ to $+105\%$
at $K=24$.

\paragraph{Experiment B: Varying target features $M$ (with $K=8$ fixed).}
Figure~\ref{fig:ablation-M} plots normalized RPS as $M$ grows from 4 to
24. Both curves degrade in near-lockstep (vanilla: 1.00 to 0.29;
rank-aware: 1.13 to 0.31); the rank-aware advantage stays in a narrow
$+7$--$+21\%$ band. As $M$ grows, target-side compute dominates, so the
context-only rank-aware savings become a smaller fraction of total work.

\subsection{Production Validation}
\label{sec:eval-production}

We deployed the DLRM-style FM+FC rank-aware variant
(\S\ref{sec:eval-comparison}) online in Taboola's recommendation platform
and compared it against the vanilla baseline. The production ranking
model runs on CPU infrastructure with autoscaled Kubernetes pods; the
autoscaler maintains a fixed latency SLA, so per-pod throughput (RPS per
pod) is the direct measure of serving efficiency.
Table~\ref{tab:production-pods} reports per-pod throughput gain and mean
latency at peak traffic for the two optimized configurations.

\begin{table}[!t]
\centering
\caption{Production deployment results for the DLRM-style ranker.
All values are percent change relative to the vanilla baseline. Per-pod
RPS is computed at peak traffic under a fixed latency SLA; p50 and p99
are mean latency.}
\label{tab:production-pods}
\begin{tabular}{lrrr}
\toprule
Variant & RPS/pod & p50 & p99 \\
\midrule
FM rank-aware        & $+25\%$   & $-17\%$ & $-26\%$ \\
FM + FC rank-aware   & $+87.5\%$ & $-30\%$ & $-33\%$ \\
\bottomrule
\end{tabular}
\end{table}

The FM optimization alone increases per-pod throughput by $25\%$. Adding
the FC optimization compounds the gain to $+87.5\%$ (equivalently, a
$47\%$ reduction in peak pod count). The FC optimization is therefore not
merely an incremental refinement over the FM optimization---it captures a
meaningful share of the remaining redundancy. This is consistent with the
theoretical analysis in \S\ref{sec:method-fc}: both stages contribute
quadratic-in-$K$ savings that compound across the forward pass.

\paragraph{Wall-clock latency.} The right two columns of
Table~\ref{tab:production-pods} report mean p50 and p99 latency. Both
percentiles reduce monotonically with more rank-aware optimization, and
p99 reduces more than p50, consistent with context-context redundancy
disproportionately affecting tail latency in the vanilla implementation.
\section{Conclusion}
\label{sec:conclusion}

We presented a rank-aware decomposition for multi-target recommendation
scoring built on a single algebraic principle: any linear or bilinear
operation over a rank-partitioned input admits an exact block
decomposition that moves context-only computation from
once-per-candidate to once-per-request. We instantiated this principle
across the dominant interaction mechanisms in modern ranking
architectures---FM pairwise products, FC projection layers, DCNv2 cross
layers, and self-attention---without changing model predictions.

Closed-form analysis and controlled ablation reveal an asymmetry:
savings scale quadratically with the number of context features and
remain insensitive to target feature count---an asymmetry not available
in pure MLP architectures, where FC-layer savings scale only linearly.
Applied to a production DLRM-style ranker without any architectural
change, the decomposition increased per-pod throughput by $87.5\%$---a
$47\%$ reduction in peak pod count at identical model predictions---while
reducing both p50 ($-30\%$) and p99 ($-33\%$) serving latency.

The identity-equivalent decomposition has a structural limit: in cross
networks and attention, each layer mixes ranks in its output, so it
applies only at the first layer. To propagate rank discipline across
the full depth of the stack, we introduced rDCN---an architectural
variant of DCNv2 that structurally forbids the off-diagonal
target-to-context coupling---and sketched an analogous variant for
self-attention. We validated rDCN offline: it matches DCNv2 accuracy
within training noise at $67\%$ fewer total FLOPs.

Two directions remain open. First, online validation: the DCN family's
small per-field context dimension makes rDCN a natural candidate for
scaling context substantially beyond what DCNv2 currently admits, and
empirical evaluation of the rank-aware attention sketch awaits a
production implementation. Second, extending rank-aware approaches to
architectures that early-fuse the target into the user-side
representation via shared self-attention. Target-attention models
(DIN~\cite{zhou2018din}, ETA~\cite{pi2020eta},
TWIN~\cite{chang2023twin}) are accommodated by the rank-aware attention
sketch of \S\ref{sec:rankaware-attn}; the harder case is self-attention
with target-token injection as in TransAct~\cite{transact2023}, where
the target's presence in the user-action sequence makes history
representations per-candidate at the first layer---breaking rank
discipline before any decomposable structure can be applied.

% ==============================================================
% Acknowledgements
% ==============================================================
\begin{acks}
I thank Shiran Schwartz and Maoz Cohen for their thoughtful feedback
on earlier drafts of this work.
\end{acks}

% ==============================================================
% Bibliography
% ==============================================================
\bibliographystyle{ACM-Reference-Format}
\bibliography{references}

\end{document}